\documentstyle[aps,pre]{revtex}          
\tighten                                  
\begin{document}                          
 \draft                                   
\twocolumn                               

\title{Diffusive Spreading
of Chainlike Molecules on Surfaces}

\author{T. Ala--Nissila$^{1-3}$, S. Herminghaus$^{4,5}$,
T. Hjelt$^{1,2}$, and P. Leiderer$^{5}$}

\address{
$^1$Research Institute for Theoretical Physics,
P.O. Box 9, FIN--00014 University of Helsinki, 
Helsinki, Finland \\
}

\address{
$^2$Laboratory of Physics, Tampere University of Technology,
FIN--33101 Tampere, Finland \\
}

\address{
$^3$Department of Physics, Brown University,
Providence, Rhode Island 02912, U.S.A.\\
}

\address{
$^4$Max--Planck Institute for Colloid and 
Interface Science, Rudower  
Chaussee 5, 12489 Berlin--Adlershof, Germany\\
}

\address{
$^5$Fakult\"at f\"ur Physik, Universit\"at Konstanz, D--7750 Konstanz,
Germany \\
}

\date{March 9, 1996}

\maketitle
\narrowtext

\begin{abstract}
We study the diffusion and submonolayer
spreading of chainlike molecules
on surfaces. Using the fluctuating bond model
we extract the collective and tracer diffusion coefficients
$D_c$ and $D_t$ with
a variety of methods. We show that
$D_c(\theta)$ has unusual behavior
as a function of the coverage $\theta$.
It first increases 
but after a maximum goes to zero as
$\theta \rightarrow 1$. We show that the increase
is due to entropic repulsion that leads
to steep density profiles for spreading droplets
seen in experiments. We also develop
an analytic model for $D_c(\theta)$ which
agrees well with the simulations.
\end{abstract}

\pacs{68.35.Fx, 68.10.Gw, 05.40.+j}

The spreading dynamics of molecularly thin oil films on solid  
substrates has gained substantial interest 
recently \cite{Hes89,Caz90,Alb92,Fra93,Haa95}. This  
is not only due to its obvious significance for technical  
applications, but also to the richness 
of the structures (layering,  
fingering) in the observed film profiles that sensitively  
depend on the molecular structure and interactions. 
As a typical feature, the development of  
molecularly thin so--called precursor films 
is observed \cite{Hes89}. The precursor film  
profile may be smooth, or  
exhibit pronounced molecular layering effects
(``dynamical layering''). 

At the  
foremost tip of the film, the flow dynamics develops into surface  
diffusion of single molecules, i.e., a 2D gas. 
In some experiments \cite{Caz90,Alb92,Fra93},
it has been found that the transition
from the dense submonolayer to the dilute film occurs much more
rapidly than expected from simple molecular diffusion.
The measured profiles are not Gaussian but assume
a steeper shape that can be well fitted by a spherical cap
\cite{Alb92} (see also Ref. \cite{Fra93}). 
An explanation for the late--time profiles has been suggested
in terms of a transition of the fluid from
a non--volatile 3D to a volatile 2D phase where the attractive
intermolecular interactions are weaker \cite{Caz90,Fra93}. 
However, it is the aim of the present study 
to provide an alternative microscopic
explanation for the steep density
profiles observed in cases where desorption of the molecules
can be neglected. By studying the
diffusive properties of chainlike molecules,
we demonstrate that such profiles can be generated
by {\it entropic repulsion alone}. 

To this end,
we have performed a systematic study of diffusion
and submonolayer spreading of athermal,
flexible chains. We concentrate on
the coverage
dependence of collective and tracer diffusion
coefficients $D_c(\theta)$ 
and $D_t(\theta)$, respectively.
In addition to being important for submonolayer
spreading dynamics, the diffusion of such molecules is
of fundamental theoretical interest.
Despite considerable experimental \cite{Gom90}
and theoretical \cite{Ala92} work on adatom
diffusion on surfaces
there are only few studies of more
complicated molecules \cite{George}.

We use the fluctuating bond (FB) model with Monte  
Carlo (MC) simulations to extract both $D_c(\theta)$ and  
$D_t(\theta)$ from the
relevant correlation functions in equilibrium. 
These results are complemented
by analysis of simulated density profiles
together with the nonlinear diffusion equation. We show that
while $D_t(\theta)$ is a decreasing function of $\theta$
as expected, 
$D_c(\theta)$ displays more complex behavior. It first
increases with $\theta$, but after a maximum value goes
to zero as $\theta \rightarrow 1$. We show that
it is this positive slope of $D_c(\theta)$ that leads to 
steep precursor profiles 
in accordance with experiments \cite{Alb92}.  
We develop a thermodynamic theory for $D_c(\theta)$ which
demonstrates that the initial increase is
due to strong entropic repulsion, which is eventually
overcome by the decreasing mobility of individual chains.
The theory gives good agreement with our numerical results.

The FB model \cite{Car88} is a 2D
lattice model of polymer chains,
where each segment excludes four nearest
and next--nearest
neighbor sites on a square lattice.
The bond lengths $b_{\ell}$ are allowed to vary 
between $2 \le b_{\ell} \le \sqrt{13}$
(in units of the lattice constant), where the upper
limit prevents bonds from crossing each other. 
The stiffness of the chains is controlled through an angle
dependent potential
$U_{\phi}=-\sum_{i=1}^n\sum_{j=1}^{N_{FB}-1}\cos(\phi)$,
where $n$ is the number of chains, $N_{FB}$ is the number of
segments in each chain and
$\phi$ the angle between two adjacent bonds.
Dynamics is introduced in the model by Metropolis
moves of single segments, with a probability of acceptance
$\min[e^{-\Delta U_{\phi}/T},1]$, where $\Delta U_{\phi}$
is the energy difference. Only such moves are allowed
that obey both the site exclusion and the bond length
restrictions. One MC time
step is defined as one attempt to move each monomer of every chain.
We note that since no
global translational modes of the chains are included, the
rigid rod
limit $T \rightarrow 0$ is not well defined in the model
\cite{Foot2}.

Simulations of diffusion were first done
for $T=\infty$ using a $180\times 180$ square lattice. We
calculated $D_t$ from the definition

\begin{equation}
D_t =
\lim_{t\rightarrow\infty}{1\over 4nt}\sum_{i=1}^n
\langle |\vec r_i(t)-\vec
r_i(0)|^2\rangle,
\end{equation}

where $\vec r_i(t)$ is the position
of $i^{\rm th}$ particle at time $t$.
The results for $N_{FB}=6$
are shown in the inset of Fig. 1. As expected, $D_t(\theta)$
is a strongly decreasing function of 
$\theta$ due to the mutual blocking of the chains.

The collective diffusion coefficient, however, shows
strikingly different behavior. It can be defined through 
the Green--Kubo relation \cite{Gom90}
$D_c = (1/2 \langle (\delta n)^2 \rangle)
\int_0^{\infty} dt \langle \vec J(0) \cdot \vec J(t) \rangle$
where $\vec J(t) = \sum_{i=1}^n \vec v_i(t)$ is the 
total diffusion current, and $\langle (\delta n)^2 \rangle$
the mean square fluctuation (in a finite area $A$).
Since the definition of $D_c$ involves cross--particle
velocity--velocity correlations, it samples the average 
collective density fluctuations of the chains instead of
just single--chain properties.
In Fig. 1 we show results for $N_{FB}=6$ conveniently 
obtained using the temporal decay of the Fourier transformed
density autocorrelation function
$S(k,t)=S_0 e^{-k^2D_C(\theta) t}$
\cite{Mak88} where $S_0$ is a constant. Great care was taken
to ensure that the hydrodynamic limit \cite{Gom90}
was reached. Initially, $D_c(\theta)$ {\it increases}
up to $\theta \approx 0.7$, after which it rapidly
approaches zero.

The initial rise of $D_c(\theta)$ is in agreement
with conclusions drawn from the experimental studies \cite{Her95}
and MC simulations of density profiles \cite{Her95,Hje94}.
We did additional studies simulating a circular
2D droplet spreading
using the FB model, and numerically solving the
nonlinear diffusion equation $\partial \theta(x,t)/\partial t =
\frac{\partial}{\partial x}
[D_c(\theta) \partial \theta(x,t)/\partial x]$ by using a 
monotonically increasing tanh fitting
function to $D_c(\theta)$ for $\theta < 0.7$. By matching
the simulated profiles to the solutions of the equation,
the form of $D_c(\theta)$ can be determined. We found it to be in 
perfect agreement with the MC results of Fig. 1. In Fig. 2
we show a comparison between
experimental \cite{Alb92}, simulated, 
and numerically calculated density profiles in the
submonolayer regime. The agreement is excellent, and these
profiles can be well fitted by a spherical cap shape with
a Gaussian foot at lowest coverages \cite{Alb92}.

Next, we studied the effects of chain length and stiffness to
collective diffusion. To study the full coverage range,
we performed careful
Boltzmann--Matano analysis of spreading profiles \cite{Gom90,Mat33}
in a rectangular geometry. The results for $N_{FB}=6$, $T=\infty$
were checked against the MC data, and very good agreement was found.
In Fig. 1 we show these 
data for $N_{FB}=24$ and 48, with $T=\infty$.
As expected, diffusion slows down but the qualitative
behavior remains the same. In Fig. 3 
we show the effect of stiffness
for $N_{FB}=6$. Again, diffusion slows down
but now also the maximum of $D_c(\theta)$ becomes less
pronounced.

To explain the unusual behavior of $D_c(\theta)$, we consider
a simple thermodynamic theory for collective diffusion
\cite{Her95}. The
strong temperature dependence of $D_c(\theta)$ evident in Fig. 2
suggests that besides the mobility of individual chains, entropy
plays an important role here. Our starting point is the
Nernst--Einstein equation \cite{Gom90}, which relates
collective diffusion to mobility $m$ via

\begin{equation}
D_c(\theta) = m \theta \biggl(
\frac{\partial \mu}{\partial\theta} \biggr),
\end{equation}

where
$\mu$ is the chemical potential. For chainlike molecules, we
approximate it by $\mu(\theta)= \mu_0 +
k_B T[ \ln(\rho) - \ln(w) ]$
where $\rho=\theta/N$ is the number of molecules with
$N$ segments per unit area, and $w$ is the number of
configurations of the chain. This gives

\begin{equation}
D_c(\theta) = m
k_B T[ 1 - \theta \frac{\partial \ln(w)}{\partial \theta}].
\end{equation}

Next, both $m$ and $w$ must be estimated as a 
function of $\theta$. These
quantities are model dependent. In general, the mobility
is expected to decrease due to mutual blocking of the chains,
and may be approximated by $m \approx m_0[1-c_n(\theta)]
\approx m_0(1-\theta^{\lambda})$,
where $m_0$ is a constant and $\lambda$ is a scaling exponent
for the probability of nearest neighbor occupation $c_n$.
Using Flory theory in 2D it can be
argued that $\lambda \approx 2.5$ \cite{Her95}.
For the entropy, in the case of a lattice polymer
with $N$ segments ($N \gg 1$) we can similarly write
$w \approx [1 + q(1-c_n)]^{N}$, where $q>1$ is a parameter.
With these approximations, $D_c$ becomes

\begin{equation}
D_c(\theta) = D_0(1-\theta^{\lambda}) \biggl( 1 +
\frac{\lambda N q \theta^{\lambda}}{1+q(1-\theta^{\lambda})}
\biggr),
\end{equation}

where $D_0$ is the single chain diffusion
constant which for single segment dynamics
scales $\propto 1/N$.
Thus, we can write $D_0=D_s/N$.
In Fig. 4 we show the behavior of this
expression as a function of coverage for various values of $N$.
The curves are
strikingly similar to the results
obtained for the FB model, and the maximum in $D_c(\theta)$
becomes relatively more pronounced for increasing $N$
\cite{Footshot}.

The theory above reveals that the underlying reason
for the unusual behavior of $D_c$ is entropic repulsion,
which at higher coverages is overcome by decreasing
mobility of single chains. For the FB model with $N_{FB}=6$,
we have numerically
extracted an effective pair interaction potential
$V_e(r)$ \cite{Ver68}
which indeed shows a strong repulsion extending up to
several lattice sites. We also find that all pair correlation
functions $G(r)$ for coverages $\theta \le 0.7$ and
$N_{FB}=6,24$ and 48 collapse to a single function which
is given by
$g(x) = G( r \theta^{\alpha}/N_{FB}^{\beta})$, where
$\alpha\approx 0.38$ and $\beta \approx 0.55$.
Finally,
it is interesting to note that
the maximum of $D_c(\theta)$ occurs at
coverages close to typical critical fractional coverages
for continuum percolation of
2D objects \cite{Pik74}.

We can use the theory derived above to make a quantitative
comparison with the FB model. To this end, we have
calculated the mobility and find that
$m \approx D_t(0)(1-\theta^{1.4})$ is
a rather good approximation \cite{tracer}.
To estimate the entropy term for segments, we
use the mean--field approximation 
$w=w_1^2w_2^{N-2}$ where $w_1$ is the entropy
arising from a segment at the end of the chain,
and $w_2$ from one in the middle of the chain.
For $N_{FB}=6$, $T=\infty$
we have numerically determined $w_1(\theta)$ and
$w_2(\theta)$, and find that they can be well fitted
by a simple tanh function \cite{Foot1}.
In Fig. 1 the solid line indicates the result
for $D_c(\theta)$ as obtained from Eq. (3)
where the only fitting parameter is
$N$. We obtain very good agreement with $N=16$ 
which is reasonable since
the true entropy is underestimated by our approximation.

To summarize, we have presented a combination of numerical
and analytic arguments to explain, how steep
submonolayer density
profiles observed in experiments \cite{Alb92}
can be obtained solely from the 
entropic diffusion and spreading 
of chainlike molecules on surfaces. For athermal molecules,
a strong entropic repulsion dominates at small and
intermediate coverages, leading to the increase of
$D_c(\theta)$.
At high coverages, the chain mobility takes over and
$D_c(\theta)$ approaches zero. The maximum of $D_c(\theta)$
is more pronounced for longer chains, but its position
is relatively weakly dependent on the details of the model.
This work also suggests that $D_c(\theta)$ can be used to
obtain information about effective chain--chain
interactions.

Acknowledgements: This work has in part been supported by
the Academy of Finland. We wish to thank S. C. Ying for
a critical reading of the manuscript.

\begin{figure}
\caption{Collective diffusion coefficient vs. coverage for the FB
model at $T=\infty$. Points are MC data for
$N_{FB}=6$, and solid line is the
theoretical fit with $N=16$. Dashed and dash--dotted lines
are Botzmann--Matano results for $N_{FB}=24$ and 48.
Inset shows the tracer
diffusion coefficient for $N_{FB}=6$. The curves have been
normalized by $D_1$ which is the diffusion coefficient for
a single segment.}
\end{figure}

\begin{figure}
\caption{Comparison between experimental density profiles of
polydimethylsiloxine spreading on silver (circles), simulations
of 2D circular droplets from the FB model with
$N_{FB}=6$ (squares), and
numerical solutions of the 1D nonlinear diffusion equation (solid
lines).}
\end{figure}

\begin{figure}
\caption{Influence of chain stiffness for collective diffusion
for $N_{FB}=6$. The data have been obtained from 
Boltzmann--Matano analysis.}
\end{figure}

\begin{figure}
\caption{$D_c(\theta)/D_s$ from Eq. (4),
with $\lambda=2.5$ and $q=6$. The solid lines from top to bottom
(at $\theta=0$)
are for $N=2,6,12,$ and 48. The maximum becomes more
pronounced for large $N$, where Eq. (4) approaches a
limiting form shown by the dashed line.}
\end{figure}


\begin{thebibliography}{100}
\bibitem{Foot2}
The present approximation for chain dynamics is valid
in the limit, where the time scale associated with
activated jumps of single
segments is much shorter than other microscopic
time scales. It has been shown in Ref. \cite{George}, that
for some molecules
the diffusion barriers scale linearly with the chain length,
which supports this approximation.
\bibitem{Footshot}
It should be noted that Eq. (4) does not give the correct
result for point--like particles, where $D_c=const.$
This limit can be built in by writing
$\mu(\theta)=\mu_0 + k_B T \ln[\theta/(1-\theta)]$
which is exact for
a Langmuir gas \cite{Gom90}, and $m \approx D_0 (1-\theta)$.
\bibitem{tracer}
For the present system, this shows that setting 
$D_t \approx m$ \cite{Gom90} is a poor approximation
and correlation effects are important.
\bibitem{Foot1}
For the FB model, fitting to
$1 + q(1-\theta^{\lambda})$ gives
unrealistic values for $q$ and $\lambda$, and the
functional form is not well satisfied.
\end{thebibliography}
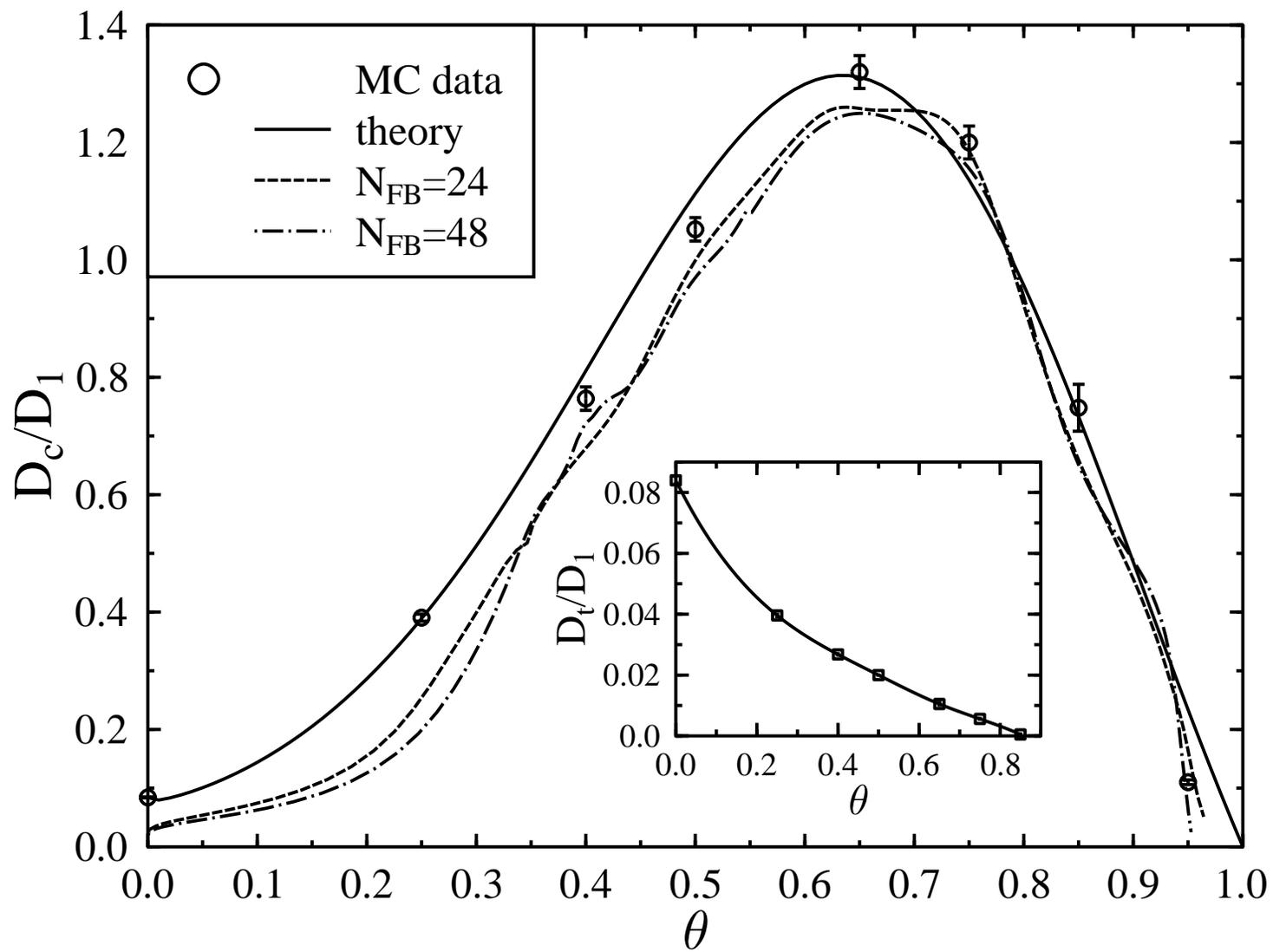
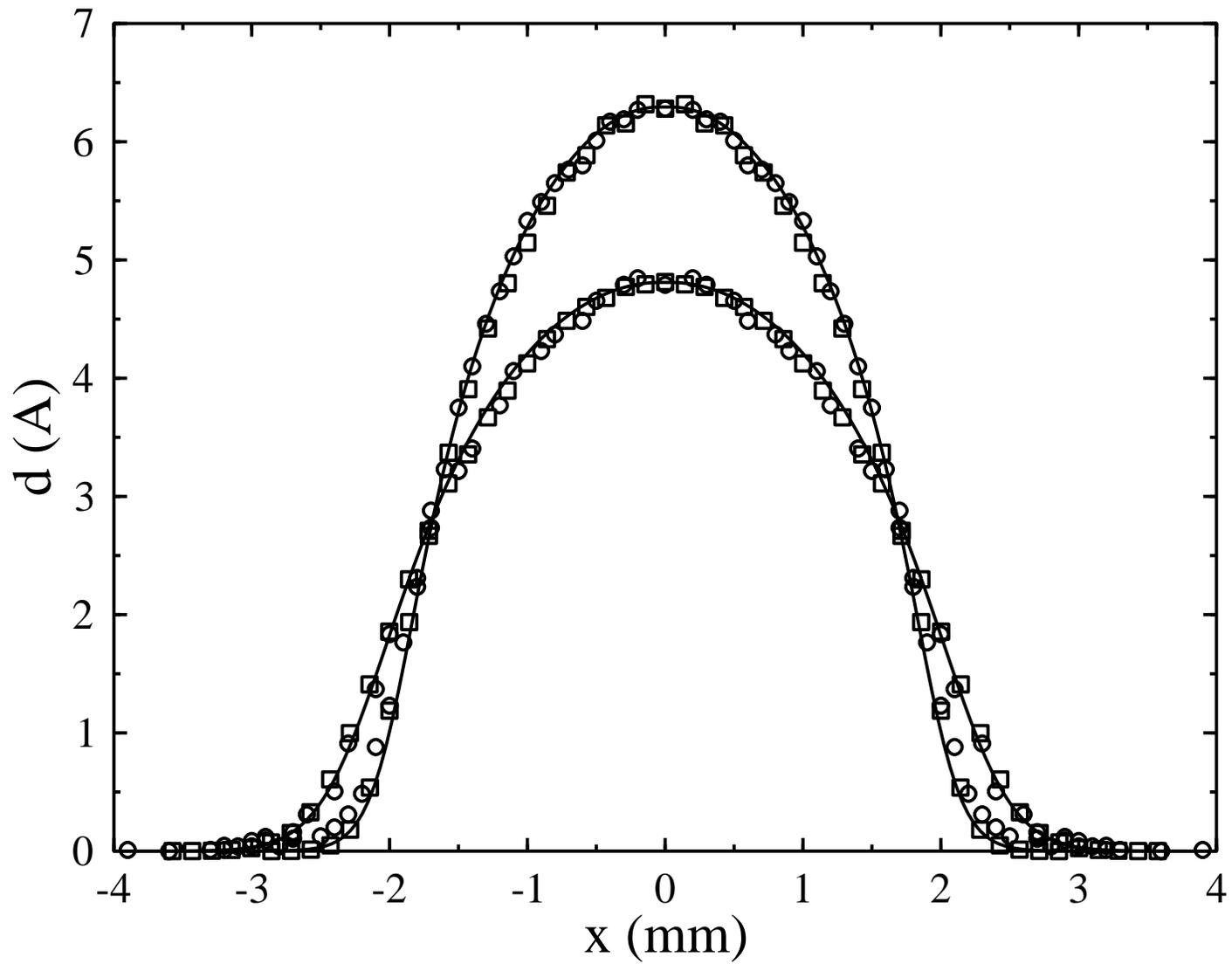
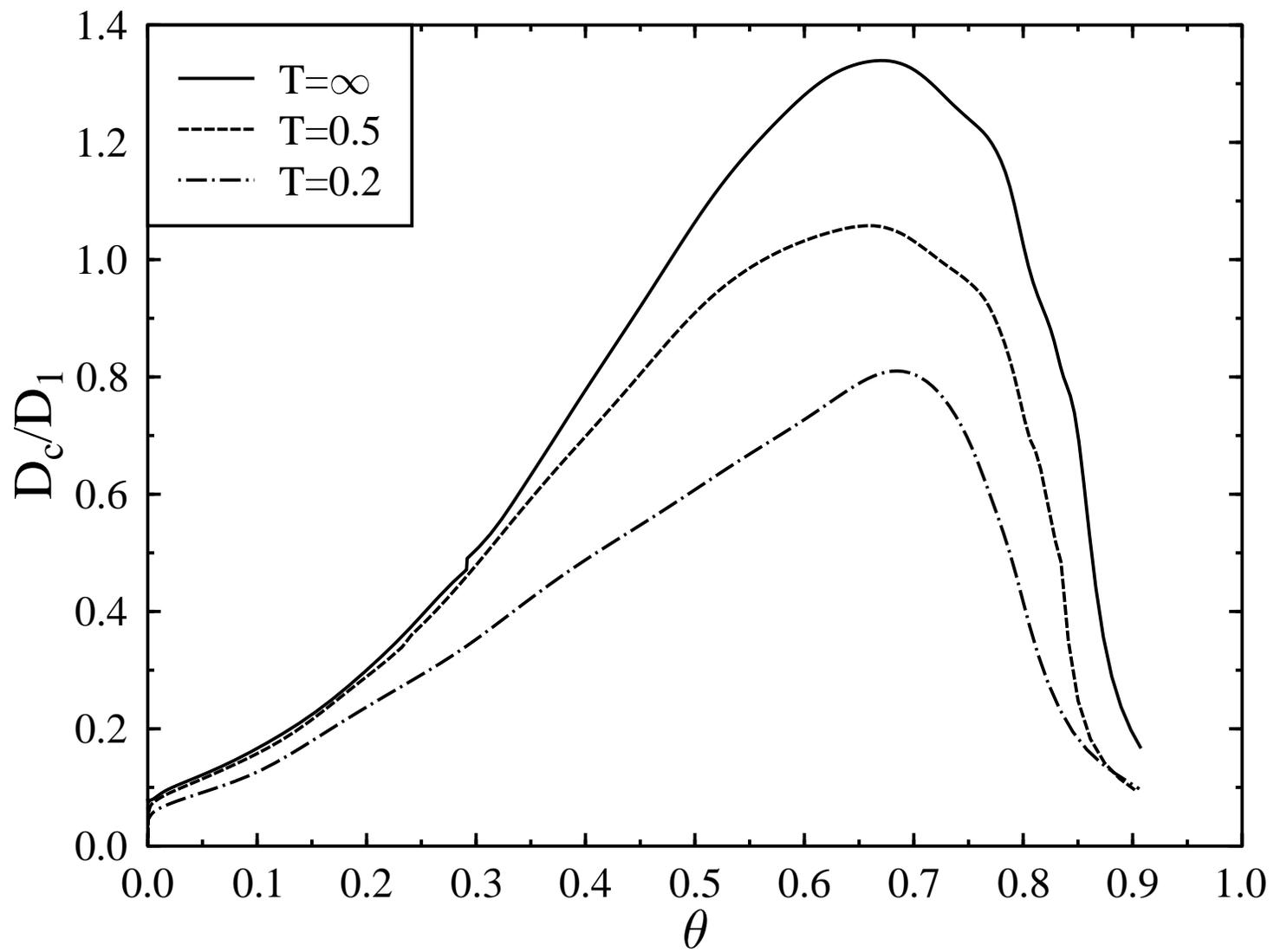
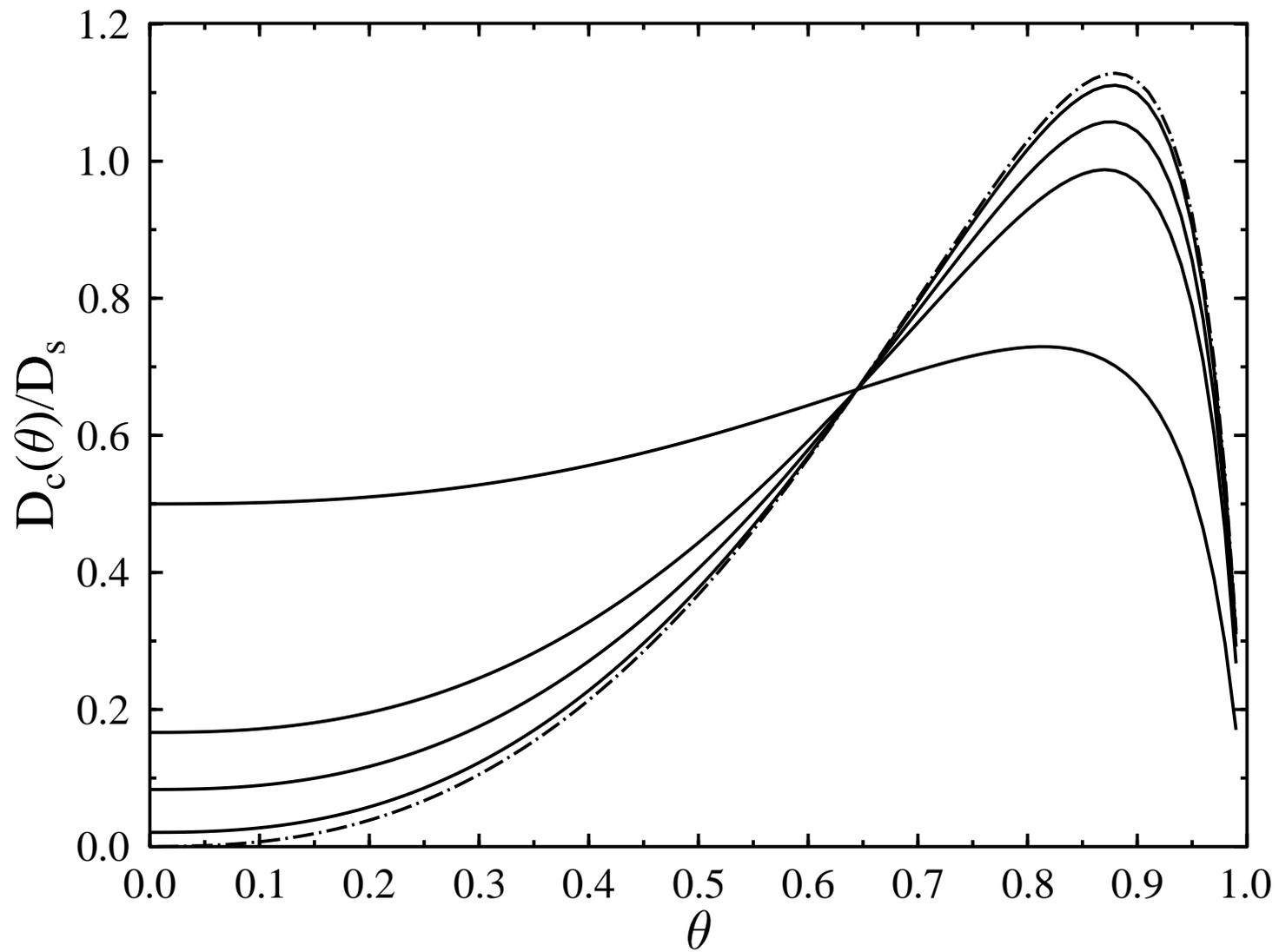
\end{document}